\documentclass[%
reprint,
superscriptaddress,
showpacs,
hidelinks,
amsmath,amssymb,
aps,
prb,
]{revtex4-1}
\usepackage{graphicx,epstopdf}
\usepackage{placeins}
\begin{document}
\title{Majorana Zero Modes in Nanowires with Combined Triangular and Hexagonal Geometry}
\author{Kristjan Ottar Klausen$^1$, Anna Sitek$^{1,2}$, Sigurdur I.\ Erlingsson$^1$ and Andrei Manolescu$^1$}

\address{Department of Engineering, Reykjavik University, Menntavegur 1, IS-101 Reykjavik, Iceland.}
\address{Department of Theoretical Physics,
	Wroclaw University of Science and Technology,
	Wybrze{\.z}e Wyspia{\'n}skiego 27, 50-370 Wroclaw, Poland.}

\vspace{10 mm}

\begin{abstract}
	The effects of geometry on the hosting of Majorana zero modes are explored in core-shell nanowires with a hexagonal core and a triangular shell, and vice versa. The energy interval separating electronic states localized in the corners from states localized on the sides of the shell is shown to be larger for a triangular nanowire with a hexagonal core, than a triangular one.  We build the topological phase diagram for both cases and compare them to earlier work on prismatic nanowires with the same core and shell geometry.  We suggest that a dual core nanowire is needed to allow for a braiding operation of Majorana zero modes at the nanowire end plane.
\end{abstract}
\vspace{-20 mm}
\maketitle
\section{INTRODUCTION}
Advances in fabrication of materials at the nanoscale have made quantum computers a realistic possibility. The quest to create quantum computers has been likened to the Space Race in the 20th-century, since encryption of government data is based on a problem that quantum computers can solve with much more ease than classical computers \cite{2018NatureRev}.

The qubit forms the basis of a quantum computer and experimentally realizing a system of multiple qubits is necessary for quantum computation. 
Numerous qubit systems have been proposed in the last two decades \cite{Bogdanov2011}. Superconducting qubits based on Josephson junction circuits have been heavily studied and are considered one of the major tracks towards quantum computation \cite{SuperQbit2020}.
 Another auspicious path is based on topological qubits \cite{Kitaev}. One of the main problems facing superconducting qubits is the scaling problem, as they are sensitive to noise which will cause accumulation of error with increased number of qubits \cite{SuperQbit2019}. Quantum computers of this type have therefore been called "Noisy intermediate scale quantum computers" \cite{SuperQbit2020}. Topological qubits have the potential to solve this issue, 
 with topological quantum computers being fault-tolerant \cite{Kitaev}.
 
A promising candidate for a topological qubit are Majorana zero modes
(MZM), also known as Majorana bound states \cite{MajoranaQubit}. They
have been predicted to have non-Abelian exchange statistics, meaning
that the order of their spatial exchange matters, making them suitable
to encode fault-tolerant quantum computation, by braiding their
world lines  \cite{AnyonComp, QmTopComp}. 

Majorana zero modes get their name from
the Italian physicist Ettore Majorana. In the year 1937 he published a
paper with a solution of the Dirac equation describing neutral particles
which are their own antiparticles,

\begin{equation}
\gamma=\gamma^{\dagger},
\label{Majorana}
\end{equation}
 thus annihilating each other on contact \cite{EttoreMajorana}. Particles with this property have since been called Majorana fermions \cite{NeutrinoMajorana}.

In superconductors, mixed states of electrons and holes can emerge as quasiparticle excitations with similar behavior \cite{TudorBok}. MZM are mid-gap excitations at zero energy, that mimic the Majorana fermion and are localized near defects or boundaries in topological superconductors \cite{Alicea_2012,Topological}.
Topological superconductors can be engineered by combining standard superconductors with semiconductors \cite{FuKane}. Currently one of the most promising platforms for realizing MZM are hybrid superconductor-semiconductor devices \cite{schapersGroup2019, Hybrid, Stanescu_2013}. When a metal or semiconductor is paired with a superconductor, the superconductivity can penetrate into the metal, making it superconducting up to a certain depth. This process of proximity-induced superconductivity is one of three necessary ingredients, along with Zeeman splitting and strong spin-orbit coupling, for engineering topological superconductors \cite{TudorBok}.
The most common model of a system with MZM is a semiconductor nanowire in 
	contact with a metal superconductor \cite{Lutchyn2010,Oreg2010}, which is a quasi-one-dimensional system. From the superconducting property, the Hamiltonian of a proximitized nanowire obtains an implicit particle-hole symmetry. With the application of an external magnetic field, time-reversal symmetry is broken and the system can support a $\mathbb{Z}_2$ topological invariant corresponding to the topological class $D$ \cite{Stanescu2011,TopoClass}.
	  The parameter space of such systems has been thoroughly explored with detailed calculations in order to predict the key conditions 
	for experimental detection \cite{Stanescu2011,Stanescu2013}.

Tubular nanowires of core-shell type provide an experimental platform to include all required aspects for hosting MZM \cite{Hybrid}. Prismatic core-shell nanowires from various semiconductors have been fabricated and continue to be an active field of research \cite{TriHex,2015TriHex,2019ProxCoreShell,Haas_2018, ZhangAruni2017,ZhangAruni2019, Sonner2019}.
Recent numerical simulations with contributions from two of the present authors have
indicated that several MZM's can be hosted in prismatic core-shell nanowires, where the 
electrons with low energies tend to localize around the prism edges \cite{Andrei,Stanescu2018}. 
This experimentally available system can be a host for previously discussed
multichain ladder models \cite{Poyhonen2014,Wakatsuki2014,Sedlmayr2016}.

In this paper we present computational results for nanowires with core geometry different from the shell geometry.  We take the cases of a triangular wire with a hexagonal core (Tri-Hex), inspired by recently fabricated structures \cite{TriHex}, and the inverse system, a hexagonal wire with a triangular core (Hex-Tri) which has also been obtained \cite{HexTri_Dick2010}. We show that the separation of energy levels of the Tri-Hex structure is significantly larger than in the case of a triangular core. With the Hex-Tri structure, the number of phase boundaries in the topological phase diagram is reduced, since the triangular core results in the ground states being localized at the three sides, rather than the six corners as is the case for a hexagonal core. The localization of MZMs is shown to correspond to the single particle localization of both configurations. The results suggest that these structures are worthy of further experimental investigation. To finalize, we discuss the possibility braiding MZM at the nanowire end planes.
\section{QUANTUM MECHANICAL MODEL AND METHODS}
In the numerical calculations we use cylindrical coordinates, with
the $z$ axis along the nanowire. 
The geometry of the wire cross-section is defined by applying appropriate boundaries to 
a discretized disc in polar coordinates $(\phi,r)$ \cite{Daday2011}.
The Hamiltonian of the cross-section describes the transverse modes of the nanowire
\begin{equation}
H_t= \frac{(p_{\phi}+eA_{\phi})^2}{2m_e} 
-\frac{\hbar^2}{2m_er} \frac{\partial}{\partial r} 
\left(r\frac{\partial}{\partial r}\right)\ ,
\end{equation}
 where $p_{\phi}=-i\frac{\hbar}{r}\frac{\partial}{\partial\phi}$ 
is the momentum in the $\hat\phi$ direction,
and $A_{\phi}=\frac{1}{2}Br$ the vector potential
associated with a magnetic field of strength $B$ oriented along the nanowire, in the symmetric gauge. The
eigenstates of $H_t$ can be written in terms of the lattice sites
\begin{equation}
|a \rangle = \sum_\kappa c_a |r_\kappa \phi_\kappa\rangle.
\end{equation}
The nanowire length is incorporated with longitudinal modes which are given by
\begin{equation}
H_l=\frac{p_z^2}{2m_e},
\end{equation}
with the corresponding eigenstates
\begin{equation}
|k \rangle =L^{-1/2}\exp(ikz) \ ,
\label{infbase}
\end{equation}
for an infinite nanowire, $L\to\infty$. In the case of a finite wire, with $z=0$ set in the middle of the nanowire,
\begin{equation}
|n \rangle =L^{-1/2}\sqrt{2} \sin\left(n\pi \left(\frac{z}{L_z}+\frac{1}{2}\right)\right) \ .
\end{equation}

The Zeeman effect due to the applied external magnetic field is given in terms of the effective Land\'{e} g-factor, Bohr magneton $\mu_B$, spin $|\sigma \rangle $ and magnetic field strength $B$,
\begin{equation}
H_Z = -g^* \mu_B \sigma B \ .
\end{equation}

In order to satisfy the Majorana property, Eq.\ \eqref{Majorana}, the system needs to be effectively 
degenerated. 
Along with the Zeeman splitting, materials such as InAs or InSb are normally used in experiments due to the possibility of obtaining strong Rashba spin-obit coupling which lifts the spin degeneracy by coupling the spin to the momentum. 

For a thin cylindrical shell the spin-orbit interaction can be modeled with the Hamiltonian
\begin{equation}
H_{SOI}= \frac{\alpha}{\hbar}( \sigma_{\phi} p_z -\sigma_z p_{\phi}) \ ,
\label{HSOI}
\end{equation}
which corresponds to the regular planar model transformed in cylindrical coordinates
\cite{Bringer2011}, where $\alpha$ is the Rashba coupling constant. 
For a prismatic shell, a more elaborated model is in principle necessary, 
that should take into account the details of the nonuniform interface between the core and 
the shell.  Recent calculations based on the 
$k\cdot p$ method indicate that the spin-orbit coupling inside the core can significantly 
increase due to the effective interface field, in the hexagonal geometry \cite{Wojcik2019}.
However, to our knowledge, no such study has been performed for the electrons situated 
in the shell. Since we shall deal mostly with electrons localized within narrow
angular regions of the shell, on the corners or on the sides, we will assume they are 
experiencing a local effective electric field in the radial direction only, and we will 
use Eq.\ \eqref{HSOI} as if that field does not depend on the angle $\phi$. This model
has been already used for describing Majorana states in core-shell nanowires with 
a simpler geometry \cite{Andrei,Stanescu2018}, and should be reasonable for a qualitative
approach.

Our model of the core-shell nanowire is based on the composite Hamiltonian
\begin{equation}
H_w= H_t + H_l + H_Z + H_{SOI} \ .
\label{wireHamiltonian}
\end{equation}
The eigenstates and energy values for a finite wire are obtained by diagonalizing the matrix
\begin{equation}
\langle an\sigma|H_w|a'n'\sigma'\rangle \ .
\end{equation}

The final ingredient needed to model MZM in the system is superconductivity. Due to the proximity effect, superconductivity is induced in the nanowire shell. The first hybrid superconductor-semiconductor nanowire systems consisted of nanowires lying on superconducting substrates \cite{DelftEvidence2012}. The proximity effect was assumed to make the whole wire superconducting given that the thickness of the nanowire is much smaller than the coherence length, $\xi$. More recently, nanowires covered with a superconductor have been fabricated with in-situ methods \cite{FluxInducedMajorana2018}. In such systems, the proximity effect can more rightly be assumed to be homogeneous in the system. However, a complete superconductor shell will invoke the Little-Parks effect which has to be taken into account in the transport through the system \cite{evenodd}.

 The current analysis essentially describes a core-shell nanowire with an insulating or hollow core and a fully proximitized semiconductor shell, which is justified in the light of experimental results of Ref.\ \cite{ABandreevXi}. Nonetheless, the authors are aware that the proximity effect deserves a more thorough treatment \cite{TudorSarma} as the temperature and sample geometry can both significantly influence the coherence length \cite{Stenuit_2004}.
 The superconducting property is incorporated by an order parameter $\Delta$ which couples two Schr\"{o}dinger equations for electrons and holes with opposite spin in the Bogoliubv-DeGennes (BdG) Hamiltonian \cite{Bogoliubov:1958km}. This has been extensively used to describe quasiparticle excitations in superconductors \cite{Jianxin},
\begin{equation}
H_{BdG}= 
\begin{pmatrix}
H_w&\Delta\\
-\Delta^*&-H^*_{w}
\end{pmatrix} \ .
\end{equation}

The eigenstates of the BdG Hamiltonian have both electron and hole components and are written in the basis 
\begin{equation}
|q\rangle =|\eta a n \sigma \rangle =|\eta g \rangle \ ,
\end{equation}
where $|g\rangle$ denotes the basis for the finite wire Hamiltonian in Eq.\ \eqref{wireHamiltonian} with the added electron-hole degree of freedom described with the isospin quantum number $\eta$. In the case of a finite wire, the longitudinal eigenstates are as in Eq.\ \eqref{infbase}. The matrix elements are obtained for $\eta=\eta'$ with
\begin{align}
\begin{split}
\langle g \eta |H_{BdG}|g'\eta' \rangle&=
	\eta[\text{Re}\langle g|H_w|g'\rangle \\&+ i\eta \langle g|H_w|g'\rangle - \mu \delta_{gg'}],
\end{split}
\label{Diag}
\end{align}
and for $\eta \neq \eta'$,
\begin{equation}
\langle g \eta |H_{BdG}|g'\eta' \rangle= \eta \sigma \delta_{\sigma,-\sigma'} \delta_{aa'} \delta_{nn'} \Delta \ .
\end{equation}
Note that the excitation energies are evaluated relative
to the chemical potential $\mu$, Eq.\ \eqref{Diag} which gives the BdG Hamiltonian an implicit particle-hole symmetry.

 By calculating the spectra of both the finite and infinite cases for increasing magnetic field strength, a closing and reopening of the quasiparticle energy gap is observed with the emergence of topological edge states in the finite length spectrum \cite{Sato_2017,HallTopo1985}.

  It has been shown that a pair of MZM will emerge for each corner in a prismatic core-shell nanowire but due to wavefunction overlap or tunneling caused by the finite width of the sides, some pairs will be shifted symmetrically above and below zero energy. 
In the case of three edges, one pair will be at precisely zero energy. The other two hybrid MZM's, slightly above and below zero energy, have been termed pseudo MZM's and will be referred to as such \cite{Andrei}. 

 Throughout this work, our chosen parameters correspond to InSb with $\gamma=\frac{1}{2} g^* m_e =0.393$, SOI parameter $\alpha=1 \text{ meV nm}$ and a superconducting gap parameter $\Delta= 0.50 \text{ meV}$.
The number of sites used to describe the nanowire cross section were between 1700-2400.

\section{SINGLE-PARTICLE ENERGIES}

\begin{figure}
	\centering
	\includegraphics[width=0.44\textwidth]{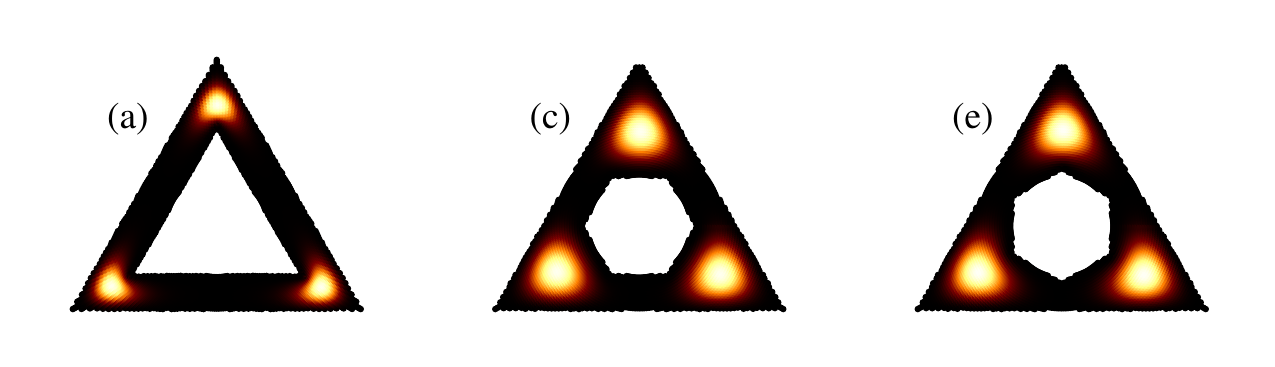}\\
	\includegraphics[width=0.44\textwidth]{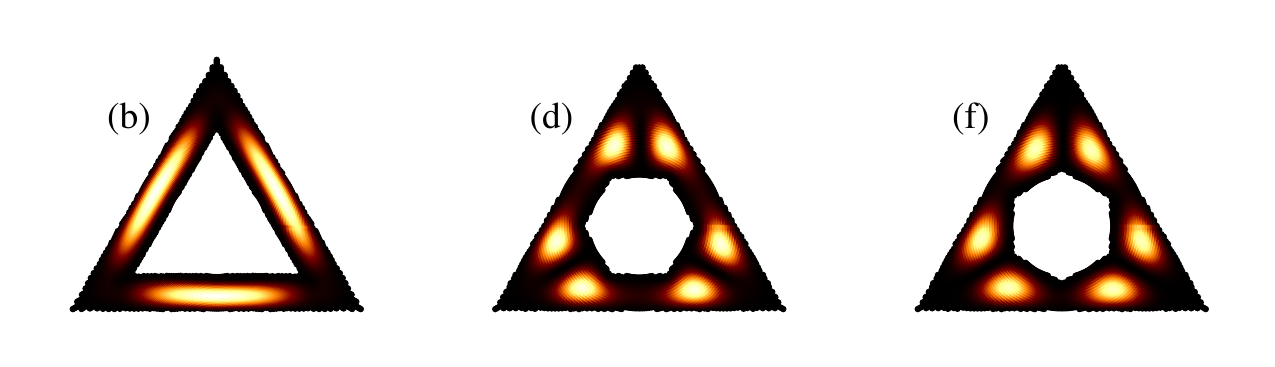}
	\caption{Single-particle cross-sectional localization of triangular nanowires with a triangular core (a,b), side-matched hexagonal core (c,d) and corner-matched hexagonal core (e,f). The upper row shows the corner localization of the first three quasi-degenerate energy states. The lower row shows the states of the adjacent energy level. The minimal shell thickness is 10 nm for the first two cases but slightly less in the corner matched case since the core is rotated with respect to the central case.}
	\label{TriLoc}
\end{figure}

Triangular nanowires with hexagonal cores have been fabricated with both side-matched (SM) cores \cite{TriHex}, Figs.\ \ref{TriLoc}(c,d) and corner-matched (CM) cores \cite{2015TriHex}, Figs.\ \ref{TriLoc}(e,f). 
To compare with earlier analysis \cite{Anna2015} we explore the single-particle localization and the energy level separation of these nanowires, in conjunction with triangular cores, Figs.\ \ref{TriLoc}(a,b).  
The hexagonal core geometry enlarges the area of the corner localized peaks, whilst the side localization is suppressed, compared to the case of the triangular core. The side states of the triangular core are split by the hexagonal core geometry, resulting in two peaks per corner. 

\begin{figure}[b]
	\centering
	\includegraphics[width=0.42\textwidth]{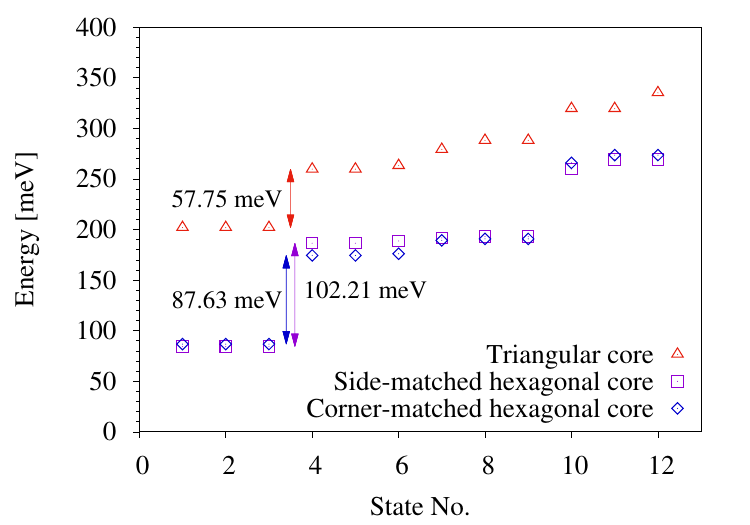} 
	\caption{Single-particle energy states for the three triangular nanowire configurations. The three lowest-energy states are nearly degenerate and localized in the corners of the shell.}
\label{TriHexEnergyt}
\end{figure}

The hexagonal core geometry further results in a larger separation between of the lowest (corner) states, Fig.\ \ref{TriHexEnergyt}, which is favourable for Majorana physics \cite{Andrei}, as it provides more robust subspace of corner states. The energy separation decreases with increasing shell thickness - as the hexagonal core gets smaller, the wavefunction overlap at the sides becomes larger.
\FloatBarrier

\begin{figure}[h!]
	\centering
	\includegraphics[width=0.45\textwidth]{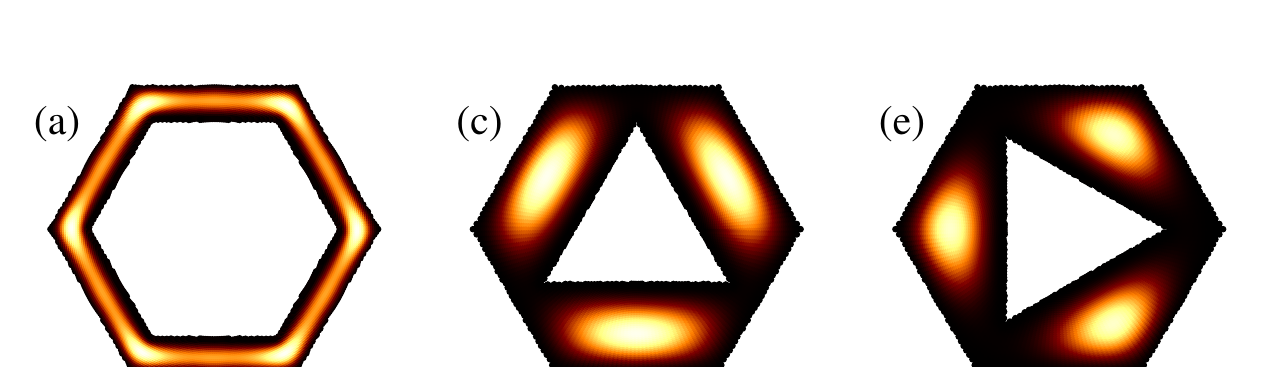}\\[0.2cm]
	\includegraphics[width=0.45\textwidth]{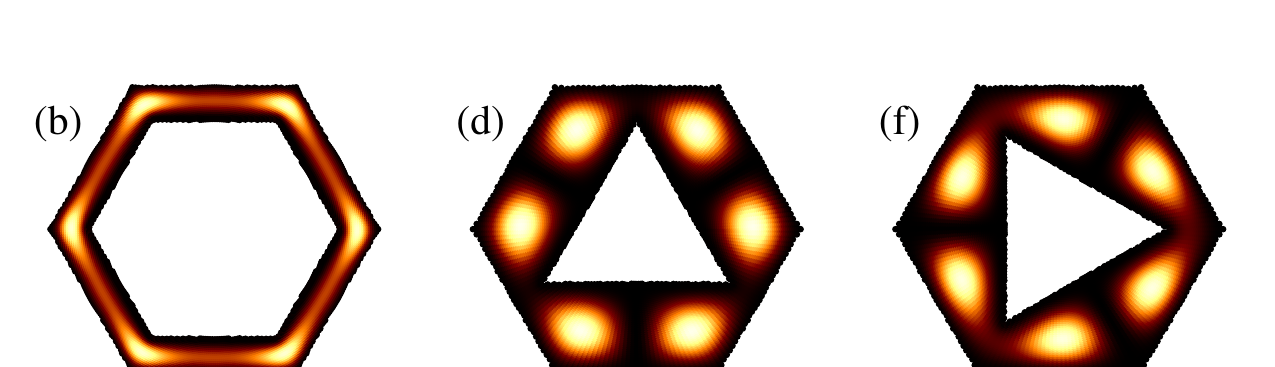}
	\caption{Single-particle cross-sectional localization of hexagonal nanowires with a hexagonal core (a,b), side-matched triangular core (c,d) and corner-matched triangular core (e,f). The upper row shows the first three quasi-degenerate energy states, which also describes the localization of the second energy level. The lower row shows the adjacent higher states. The minimal shell thickness is 10 nm for the first two cases but slightly more in the corner matched case since the core is only rotated with respect to the central case.}
\label{HexTriLoc}
\end{figure}

\begin{figure}[h!]
	\centering
	\includegraphics[width=0.45\textwidth]{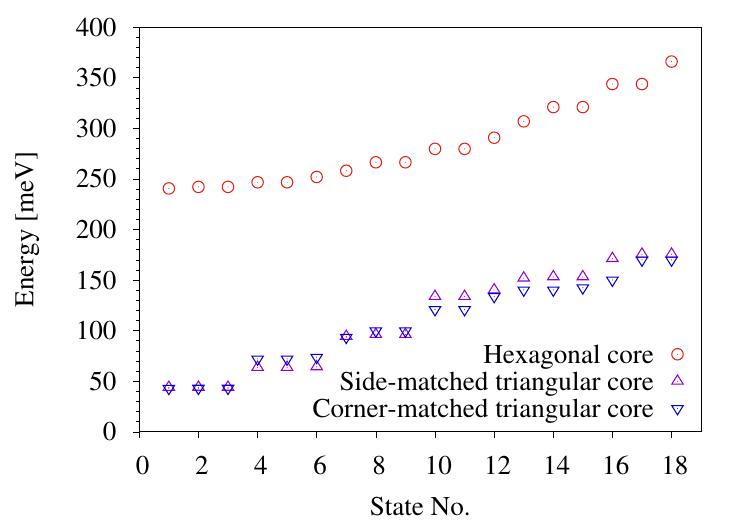} 
	\caption{Single-particle energy states for the three hexagonal nanowire configurations.}
	\label{HexTriEnergyt}
\end{figure}

In the case of a hexagonal wire, Fig.\ \ref{HexTriLoc}, the triangular core outline three sides with an enlarged area, compared to the hexagonal core. To the best of our knowledge, the corner-matched configuration (e,f) has not yet been fabricated, but is included in the analysis for the sake of completeness. The separation of energy levels is larger for the case of a hexagonal wire with a triangular core compared to a hexagonal core, Fig.\ \ref{HexTriEnergyt} but not as large as in the case of triangular wires. This is due to the overlap between localization peaks, which is clearly visible in Fig.\ \ref{HexTriLoc}(f).
 We say that the three states are quasi-degenerate as the degeneracy pattern is 1-2, 2-1 which becomes more evident for higher states in both Fig.\ \ref{TriHexEnergyt} and Fig.\ \ref{HexTriEnergyt}.
\FloatBarrier

\FloatBarrier
\section{MAJORANA ZERO MODES AND TOPOLOGICAL PHASE DIAGRAMS}
\begin{figure}[b]
	\centering
	\includegraphics[width=0.5\textwidth]{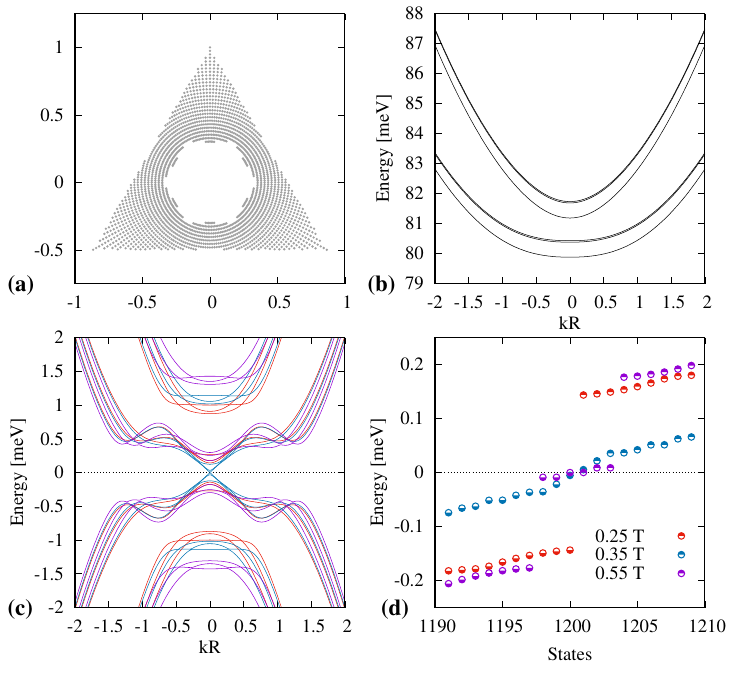}
	\caption{(a) Discretized nanowire cross section with Tri-Hex geometry, radius of 50 nm and 10 nm minimal shell thickness. \newline (b) Corresponding energy dispersion for an infinite wire in the presence of an external magnetic field of magnitude $0.55$ T. \newline (c) BdG quasiparticle spectra for an infinite wire showing the closing of the energy gap as the longitudinal external magnetic field strength is increased. (d) Corresponding energy spectra of a finite wire of length 200 R, showing the emergence of Majorana Zero Modes. The line colors in figure (c) correspond to the magnetic field values shown in figure (d).}
	\label{BdGspectra}
\end{figure}

First we show a set of results for a nanowire with the 
	cross section illustrated in Fig.\ \ref{BdGspectra}(a).  The radius of the 
	nanowire i.e.\ the distance between the center and the external corners, is $R=50$ nm,
	and the side thickness is 10 nm. In Panel (b) we show the energy dispersion in
	such a nanowire with infinite length, vs.\ the longitudinal wavevector times the nanowire
	radius, showing the Zeeman splitting due to the external magnetic field applied parallel to the nanowire. Next, in 
	Figs.\ \ref{BdGspectra}(c,d), we show the energy spectra for the BdG Hamiltonian,
	for the the nanowire of infinite and finite length respectively. In the later case the 
	length was 200 R. Three values of the longitudinal magnetic field are shown, to indicate
	the onset of the MZM. Whereas in Panel (c) we can only observe the vanishing 
	gap (at $k=0$), in Panel (d)  we have the complementary information on the 
	six states created around zero energy, one for each corner of the nanowire section,
	times two due to the electron-hole symmetry.

By calculating the BdG spectra for a range of magnetic field strengths and values of the chemical potential, we can find for which set of parameters the BdG spectra closes. The closing of the BdG energy gap signifies a phase transition of the system from a topologically trivial state to a non-trivial one, with the emergence of a pair of edge states. The system can not be adiabatically deformed back to the original trivial state, just as a sphere can not be continuously deformed into a doughnut, therefore the states are considered topologically distinct \cite{TopoPhase_Shou-Cheng,AdiaTopoPhase2013}. 
 By plotting the ratio of the BdG energy and the superconducting gap parameter $\Delta$ at $k=0$, as a function of the magnetic field strength and the chemical potential, a phase diagram is obtained \cite{Stanescu2011,AkhmerovTopoPhase}. Since there are three states at the lowest energy, for both the Tri-Hex and Hex-Tri case, the system can host three pairs of edge states and there will be a closing of the BdG spectra for the emergence of each one. We will therefore have three curves on our phase diagrams which are boundaries between phases with different number of pairs of edge states  \cite{TopoPhase2017,Exp_phase_diag_Chene1701476}.
Topological phase diagrams of prismatic core-shell nanowires with uniform geometry have been recently studied  \cite{Stanescu2018}. Here we explore the effect of the core geometry on the phase diagrams. Since a realistic nanowire is not strictly symmetric, from now on we choose to slightly break the cross-sectional symmetry of our nanowires with a weak transverse electric field. 

\subsection{Tri-Hex nanowire geometry}

\begin{figure}[b]
	\centering
	\includegraphics[width=0.42\textwidth]{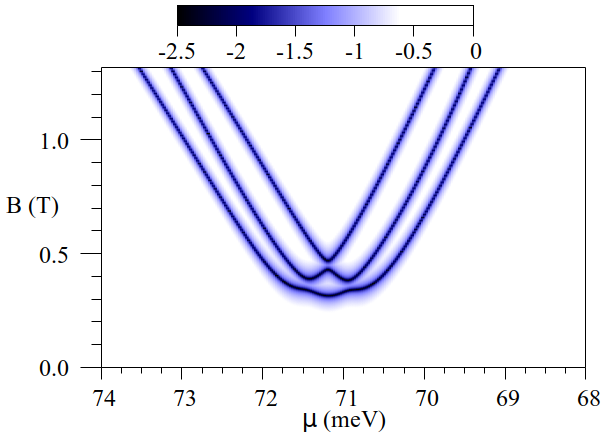}
	\caption{Phase diagram for a side-matched Tri-Hex structure with 12.5 nm shell thickness in a weak transverse electric field. The colourscale represents the ratio of the energy to the superconducting gap parameter $\Delta$ at $k=0$ on a $\log_{10}$ scale.}
	\label{TopoTriHex125nm}
\end{figure}

For a minimal shell thickness of 10 nm, the phase diagram of the Tri-Hex systems consists of three lateral parabolas, which is the same phase diagram as for three isolated one dimensional nanowires \cite{Stanescu2018}. In order to observe signs of wavefunction overlap between corners, the minimal shell thickness is increased to 12.5 nm, correspondingly the chemical potential is shifted to lower values compared to Fig.\ \ref{BdGspectra}.
Contrary to the phase diagram of three isolated wires, the tunneling removes the crossing of the phase boundaries, Fig.\ \ref{TopoTriHex125nm}.
 The same phase diagram is observed for all three configurations of core geometry so only the side-matched case is presented.
\FloatBarrier 
We conclude that the effect of the hexagonal core is minimal for shell thickness up to 12.5 nm. Larger values will decrease the separation of energy levels, which is undesirable for hosting MZM \cite{Andrei} and are thus not further elaborated on here. 
\subsection{Hex-Tri nanowire geometry}
\FloatBarrier

\begin{figure}[h!]
	\centering
	\includegraphics[width=0.45\textwidth]{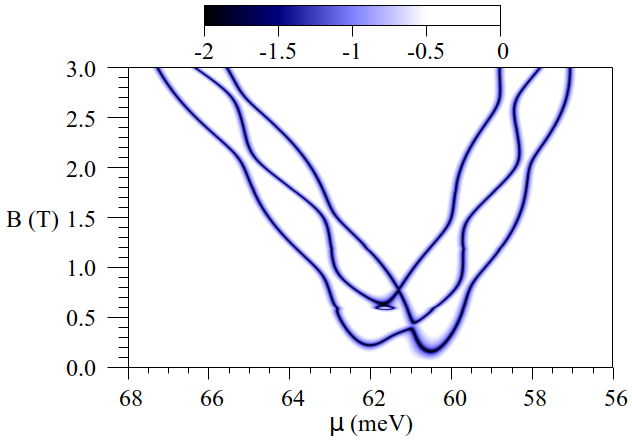}
	\caption{Phase diagram for a side-matched Hex-Tri structure with 10 nm shell thickness. Three topological phases, corresponding to the three sides of the triangular core are distinguished.
}
	\label{Pd_HexTri_SiM_B01p5}
\end{figure}

For a hexagonal wire with a triangular core, we can expect to see more signs of wavefunction overlap between the sides of the wire, as the separation of energy levels is much smaller than in the case of the triangular wire, Fig.\ \ref{HexTriEnergyt}. As the differences between the side-matched and corner-matched cases is small, the phase diagram for the side-matched structure is presented only.
In the topological phase diagram the presence of a threefold phase boundary signifies that the particles form three channels induced by the triangular core geometry, Fig.\ \ref{Pd_HexTri_SiM_B01p5}. The topological phase is entered at a lower value of both the chemical potential and magnetic field strength, compared to the Tri-Hex case, Fig.\ \ref{TopoTriHex125nm}.
As shown in Ref.\ \cite{Andrei}, orbital Zeeman effects both skew the phase boundaries and lower the magnetic field strength threshold for the topological phase. Furthermore, combined with the spin Zeeman effect, they are responsible for the curious elliptic island, Fig.\ \ref{Pd_HexTri_SiM_B01p5}, around $\mu=61.5 \text{ meV}$ and $B=0.6 \text{ T}$. 

\begin{figure}
	\centering
	\includegraphics[width=0.4\textwidth]{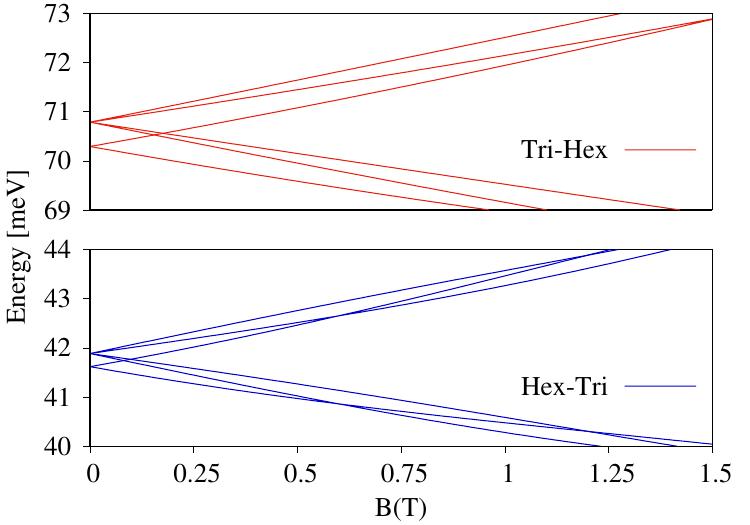}
	\caption{Oscillations of the lowest energy transverse
		states with increasing the magnetic field, compared for a side-matched Tri-Hex and Hex-Tri wire geometries.}
	\label{ZeemanOsc}
\end{figure}

In Fig.\ \ref{ZeemanOsc}, the lowest energy transverse states are shown to oscillate with increasing strength of the external magnetic field due to the orbital Zeeman effect. The oscillations are more pronounced for the Hex-Tri geometry, which is echoed in the corresponding phase diagram. The phase boundaries are three in number but a hexagonal core results in six phase boundaries \cite{Andrei}. The effect of the triangular core geometry is therefore quite significant in this case, mainly in that the number of phase boundaries is halved, compared to a hexagonal core.
\FloatBarrier
The possibility of gap closing at non-zero values of the wave vector in the continuous spectrum was also considered. In all of the studied cases, the same phase diagram was obtained, meaning that these topological phases are stable \cite{Andrei}.
\FloatBarrier
\section{TRANSVERSE AND LONGITUDINAL MAJORANA ZERO MODE LOCALIZATION}
\begin{figure}
	\centering
	\includegraphics[width=0.425\textwidth]{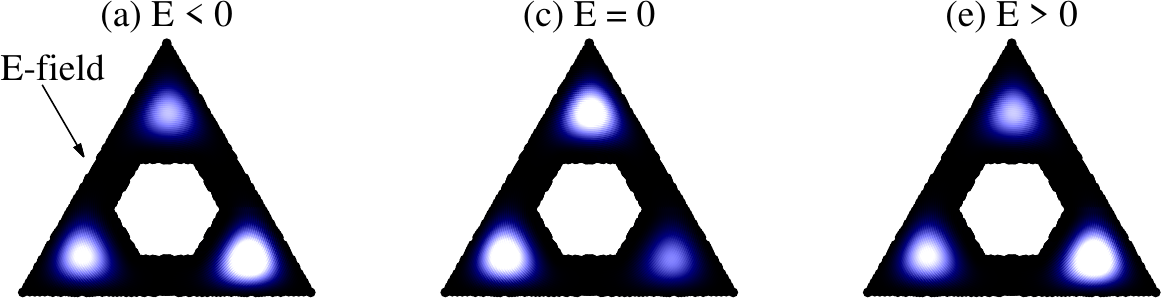}\\[0.25cm]
		\includegraphics[width=0.425\textwidth]{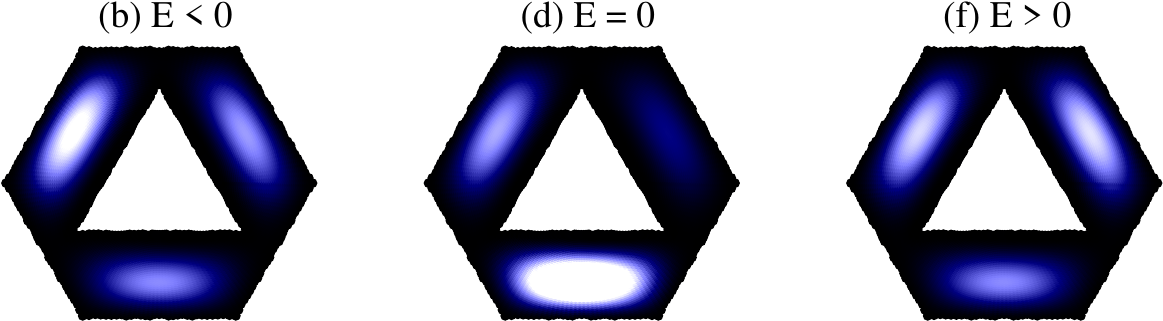}
	\caption{ Localization of Majorana zero modes at the nanowire ends for side matched Tri-Hex and Hex-Tri structures. The arrow shows the directionality of the applied transverse electric field for all instances, that breaks the localization symmetry.}
	\label{TriHex_Maj}
\end{figure}

 To explore the correspondence between the single-particle localization and MZM localization, the BdG probability density of the nanowire cross section at the end of the wire is calculated.  
We find that for both the Tri-Hex and Hex-Tri structures, the localization of MZM's, Fig.\ \ref{TriHex_Maj}, coincides with the single-particle localization, Figs.\ \ref{TriLoc} and \ref{HexTriLoc}, in that we have states localized at the largest area. The ideal MZM Fig.\ \ref{TriHex_Maj}(c,d) at $E=0$ differ from the pseudo MZM (a,b,e,f) by different weights of the lowest (corner) states, controllable with the electric field. The degeneracy pattern 1-2, 2-1 is reflected in the localization.

 Only the side-matched cases are presented, as the difference to the corner-matched cases is negligible.

\begin{figure}[h!]
	\centering
	\includegraphics[width=0.49\textwidth]{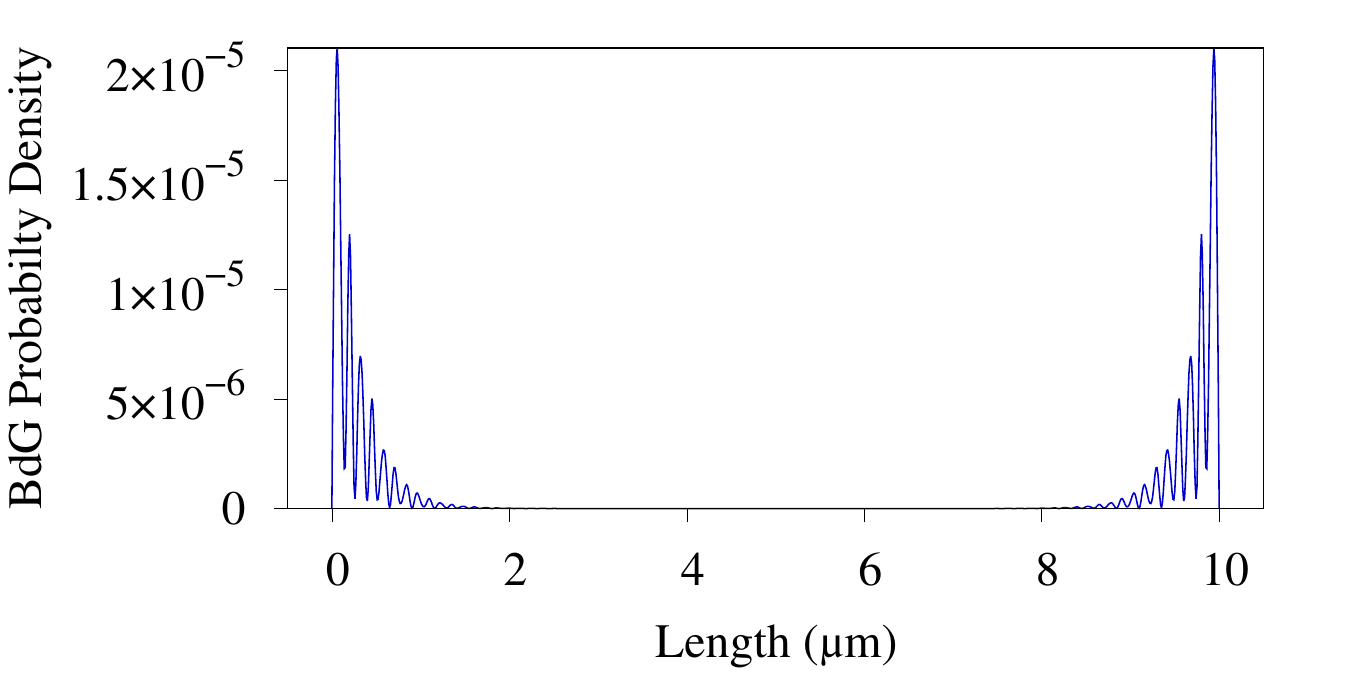}
	\caption{Single site lengthwise localization of an MZM. The probability density is greatest at the nanowire ends and falls off exponentially towards the nanowire center.}
	\label{Longiloc}
\end{figure}

To confirm that the MZM are localized at the ends of the nanowire, the BdG probability density of the corresponding state is calculated as a function of the nanowire length, for a given point on the nanowire cross section.
The longitudinal localization for the corresponding top corner of the side-matched Tri-Hex, Fig.\ \ref{TriHex_Maj} can be seen in Fig.\ \ref{Longiloc}. As expected, we observe strong edge localization which is characteristic of the MZM and topological edge states in general \cite{FuKane, Liu_2017}.
\FloatBarrier

\section{FUTURE OUTLOOK}
In the light of the zero bias conductance peaks not being conclusive evidence \cite{Zhang2019}, there is now a general consensus that braiding or fusing MZM is required as conclusive evidence to confirm experimentally their realization \cite{Zhang2019, 2018NatureRev}.  Solely demonstrating topological phases as we have done here, is only the necessary foundation. Various schemes for braiding have been proposed \cite{BraidingBeenakker}. In particular, braiding protocols for Y-junctions have been heavily studied \cite{BraidYjuncError,BraidError2}. In such a system, one of the emerging problems is the misalignment of the longitudinal magnetic field, as the Y-junction does not lie on a single axis \cite{BraidingYjunctionTuningFork}. Braiding MZM's in the plane of the nanowire end will circumvent this problem. However, this can not be done in a single core nanowire system as the MZM's would always meet up along the shell and annihilate. 

\begin{figure}[h!]
	\centering
\includegraphics[width=0.28\textwidth]{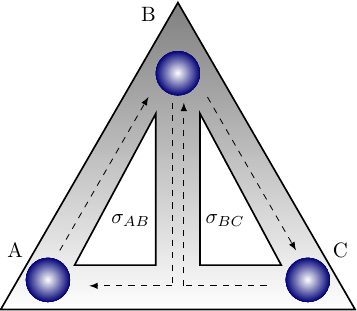}
\caption{End plane of a hypothetical dual core-shell nanowire. The genus 2 surface provides sufficient motional degrees of freedom for the braiding operations $\sigma_{AB},\sigma_{BC}$.}
\label{2core}
\end{figure}

By fabricating a core-shell nanowire with two insulating cores or effectively so via doping, Fig.\ \ref{2core}, this problem can be overcome.
The next problem is then how the MZM's can be moved around in the nanowire end plane. Even though the MZM can be manipulated with an external electric field, that may not provide sufficient control to perform a braiding operation.
Lastly, a readout is needed, based on a coupling of the fermion parity to an observable \cite{BraidingBeenakker} in order to confirm that $\sigma_{AB}\sigma_{BC}\neq\sigma_{BC}\sigma_{AB}$. 
\FloatBarrier
Formulation of an exact braiding scheme is beyond the scope of this article. Here, we only intend to stimulate the discussion about braiding in the nanowire end planes, as the nanowire edges model three strands, needed to demonstrate the simplest non-commutative braiding operation.

\section{SUMMARY AND CONCLUSIONS}
By combining hexagonal and triangular geometry in nanowires of core-shell type, both energy levels and topological phases can be influenced.  With the polygonal geometry, a well separated group of corner states is obtained. Here we show that
the separation of corner-localized energy levels is much larger for triangular nanowires with hexagonal cores, compared to those with triangular cores of the same minimal shell thickness, whilst the effect on the topological phase is minimal. On the contrary, for the complimentary configuration of a hexagonal wire with a triangular core, the effect on energy level separation is minimal whilst the effect on the topological phase boundaries is significant. We find for both configurations that the localization of the Majorana zero modes coincides with the single-particle localization. The problem of braiding Majorana zero modes in a nanowire end plane is addressed and a split/dual core structure is proposed as a system with the necessary motional degrees of freedom for a three strand braiding operation. 
\begin{acknowledgments}
This research is supported by the Reykjavik University Research Fund, project no.\ 218043 and the Icelandic Research Fund, grant no. 206568-051. We are grateful to Kristof Moors and Pujitha Perla for fruitful discussion about split/dual core-shell nanowires and braiding. 
\end{acknowledgments}
\bibliographystyle{apsrev4-1}
\bibliography{CSNreferences_v06}
\end{document}